\documentclass[aip,twocolumn,groupedaddress,reprint]{revtex4-1}
\usepackage{amsmath,amsfonts,amssymb,dcolumn,graphicx}
\usepackage{lineno}
\usepackage{xr}
\usepackage{floatrow}
\usepackage{float}

\floatstyle{plaintop}
\restylefloat{table}

\begin{document}

\def\bra#1{\langle #1|}
\def\brak#1{\langle #1}
\def\ket#1{|#1\rangle}
\def\braket#1#2{\langle #1|#2\rangle}
\def\expec<#1|#2|#3>{\langle #1|#2|#3\rangle}
\def\BRA#1{\Big\langle #1 \Big|}
\def\KET#1{\Big|#1\Big\rangle}
\def\EXPEC<#1|#2|#3>{\BRA{#1}#2\KET{#3}}
\def\half{{\textstyle \frac{1}{2}}}
\def\quart{{\textstyle \frac{1}{4}}}
\def\sixth{{\textstyle \frac{1}{6}}}
\def\eighth{{\textstyle \frac{1}{8}}}
\def\ddt#1{{\textstyle \frac{\partial #1}{\partial t}}}
\def\rezi#1{{\textstyle\frac{1}{#1}}}
\newcommand{\com}[2]{\big[ #1, #2 \big]}
\newcommand{\doublecom}[3]{\Big[ \big[ #1,#2 \big] , #3 \Big]}
\newcommand{\bigcom}[2]{\Big[ #1, #2 \Big]}
\newcommand{\hambar}{e^{-\ha\usepackage{subcaption}t{T}}\hat H e^{\hat{T}}}
\newcommand{\tento}[2]{$#1\!\cdot\! 10^{#2}$}

\newcommand{\Nexc}{N_\mathrm{rank}}
\newcommand{\Ncom}{N_\mathrm{com.}}
\newcommand{\Nact}{N_\mathrm{act.}}
\newcommand{\Nfac}{N_\mathrm{fac.}}
\newcommand{\cs}{$\times$}
\newcommand{\st}{$\star$}
\newcommand{\bx}{$\odot$}

\newcommand{\be}{\begin{equation}}
\newcommand{\ee}{\end{equation}}

\renewcommand{\epsilon}{\varepsilon}
\newcommand{\dott}{\dot{T}}
\newcommand{\excomega}{\omega}

\title{Response Formalism within Full Configuration Interaction Quantum Monte Carlo: Static Properties and Electrical Response}
       


\author{Pradipta Kumar Samanta}
\affiliation{
  Institut f\"ur Theoretische Chemie, Universit\"at Stuttgart,
  D-70569 Stuttgart, Germany
}
\affiliation{
  Max Planck Institute for Solid State Research ,
  D-70569 Stuttgart, Germany
}
\author{Nick S. Blunt}
\affiliation{
University of Cambridge, Lensfield Road, Cambridge CB2 1EW, U.K.
}
\author{George H. Booth}
\email{george.booth@kcl.ac.uk}
\affiliation{
 Department of Physics, King's College London, Strand, London WC2R 2LS, U.K.
}

\date{\protect{\em \today}}

\begin{abstract}
We formulate a general, arbitrary-order stochastic response formalism within the Full Configuration Interaction Quantum Monte Carlo framework. This modified stochastic dynamic allows for the exact response properties of correlated multireference electronic systems to be systematically converged upon for systems far out of reach of traditional exact treatments. This requires a simultaneous coupled evolution of a response state alongside the zeroth-order state, which is shown to be stable, non-transient and unbiased. We demonstrate this with application to the static dipole polarizability of molecular systems, and in doing so, resolve a discrepancy between restricted and unrestricted high-level coupled-cluster linear response results which were the high-accuracy benchmark in the literature.
\end{abstract}

\maketitle

\section{Introduction}
\label{sec:introduction}

The change in a system due to a perturbation is at the heart of experimental techniques to probe a range of properties. Whilst energies are certainly of interest, most properties of relevance can be defined as the response of a wavefunction or its expectation value to a perturbative change in the Hamiltonian defining the system. These cover responses due to applied fields, changes in the environment, or perturbative coupling of the Hamiltonian to a nuclear spin or neglected effects such as spin-orbit coupling. Rather than computing this response as a finite difference of properties between the perturbed and unperturbed system, response theory has provided a powerful framework to directly compute these quantities, defined as derivatives of expectation values of the original Hamiltonian due to the perturbation\cite{Monkhorst1977,Dalgaard1983,Christiansen:1998p36,Helgaker2012}. The success of this approach has greatly expanded the utility and scope of \emph{ab initio} quantum chemistry, and allowed methods such as coupled-cluster and other correlated methods to be applied to problems ranging from the calculation of nuclear magnetic shielding constants to circular dichromism, and much more between\cite{GaussNMR1995,GaussGTensor2009,Dorando2009,Crawford17}. This broad scope of application is now seen as essential to the success of the field.

In contrast to quantum chemistry, projector quantum Monte Carlo methods generally have had a more limited availability of response quantities, due mainly to the fact that the perturbations do not in general commute with the unperturbed Hamiltonian. This fact necessitates the computation of pure expectation values, for which projector quantum Monte Carlo techniques are not in general well suited\cite{Foulkes2001,Rothstein15}. Despite this, the computation of response properties has been performed successfully within diffusion quantum Monte Carlo via numerical finite difference\cite{Moroni1992,Moroni1995}, or by integrating along a path in a Berry-phase formalism\cite{Umari2005,Umari2009}. Further advances in reptation and forward-walking algorithms in QMC are now beginning to be able to explore pure expectation values and response functions\cite{Baroni1999a,Wagner2007,Gaudoin07,Rothstein11,Coccia2012a,Per12,Rothstein15,Motta2017}. However, it is clear that these are exceptionally difficult quantities to compute in general for these methods.

In this paper, we outline an approach for the direct and stable computation of arbitrary order response properties within the framework of Full Configuration Interaction Quantum Monte Carlo method (FCIQMC), a projector QMC approach. In contrast to related projector QMC methods, this approach is constructed in a discrete basis of antisymmetric, orthogonal Slater determinants\cite{Booth2009,Cleland2010,Booth11,Booth2013}. This approach for response properties inherits the advantages of the ground and excited state FCIQMC approach in being systematically improvable to exactness (equivalent to FCI), in a non-transient fashion.
The sign problem in this approach is suppressed via a combination of annihilation events between oppositely signed walkers which sample the wavefunction amplitudes, along with the initiator approximation which aids in promoting annihilation events relative to spawning\cite{Kolodrubetz2013,Spencer2012}.

Following the use of FCIQMC in calculating energies of chemical and solid-state systems, this development allows for the routine calculation of arbitrary analytic energy derivatives of interest within the same framework, in a similarly systematically improvable fashion. This is achieved via the simultaneous sampling of multiple walker distributions, corresponding to these derivative wavefunctions. These can then be used to accumulate unbiased statistical estimates of response functions of arbitrary order, which can be averaged over time. 
Dynamic response functions have been considered previously within FCIQMC frameworks\cite{Booth2012,Blunt2015b}, however those algorithms have always led to transient estimates of these response functions, which are avoided in the stable sampling detailed in the following construction. In this work, we restrict ourselves to static responses of the ground state of chemical systems, and to only linear, first-order response functions. However, we will detail a more general procedure to obtain arbitrary-order response, and future work will investigate the performance of the approach for higher-order response, including the response of excited states, application to solid-state systems, and extension to a stable, non-transient sampling of dynamic response functions.


This work builds on a number of recent developments within the FCIQMC framework. The first is an excited-state approach, where multiple walker distributions stochastically evolve to sample each state of interest simultaneously\cite{Blunt2015a}. This was achieved with a stochastic orthogonalization procedure between the statistical sampling of the different low-energy states.
A key further development has been the efficient sampling of reduced density matrices (RDM), and their use in the calculation of \emph{pure} (symmetric) expectation values and properties of chemical systems\cite{Booth2012a,Overy2014}. This contrasts with a projected estimate which is generally used in the computation of the energy, but can only be used for operators which commute with the Hamiltonian being sampled. A central concept to allow for this was to ensure that two, independent walker distributions were used to sample the wavefunction of each state, so that unbiased expectation values could be obtained.
These unbiased RDMs were shown to be able to compute essentially exact nuclear forces and dipole moments as expectation values, while the polarizability was also computed by using a finite field difference approach\cite{Thomas2015}. These RDMs have also recently been extended to compute expectation values between any two sampled states via the construction of distributed-memory transition density matrices\cite{Blunt2017b}, and much of this infrastructure will also be used within this work. 


In section~\ref{sec:background}, we start by providing a brief overview of the FCIQMC method, with details concerning the sampling of transition density matrices between two simultaneously evolving walker distributions representing different sampled states of the system. 
In section~\ref{sec:theory}, we derive a modified set of coupled master equations for the stochastic dynamics within FCIQMC, to ensure that we can sample arbitrary responses of walker distributions to a perturbation. 
We further provide remarks on the implementation involving the solution to the response equations and subsequent computation of the second order properties through sampling respective transition RDMs.
Finally, in section~\ref{sec:results} we present results for different components of the dipole polarizability for chemical systems, demonstrating the applicability of our method, and investigate the convergence of both systematic and random errors in the calculations.

\section{Theoretical Background}
\label{sec:background}

\subsection{The FCIQMC Method}

We provide here a brief overview of the underlying framework of Full Configuration Interaction Quantum Monte Carlo (FCIQMC) for sampling ground-state wavefunctions\cite{Booth2009,Booth2010,Booth11,Cleland2010}.
FCIQMC is a stochastic method to solve the imaginary-time Schr\"odinger equation, which has the form:
\begin{equation}
\frac{\partial}{\partial\tau}\ket{\Psi} = - \hat{H}\ket{\Psi}.
\label{eq:tdse}
\end{equation}
Starting with an initial state $\ket{\Psi(0)}$, which is not orthogonal to the true ground state, the propagation of Eq.~\ref{eq:tdse} leads to a dynamic given by
\begin{equation}
\ket{\Psi(\tau)} =  e^{-\tau\hat{H}}\ket{\Psi(0)},	
\label{eq:propagate}
\end{equation}
which in the long-time limit, converges to a solution proportional to the ground state of the electronic system, $\ket{\Psi_0}$,
\begin{equation}
\ket{\Psi_0} \propto \lim_{\tau\rightarrow \infty} e^{-\tau\hat{H}}\ket{\Psi(0)}.
\end{equation}
The numerical integration of the time-dependent Schr\"odinger equation in Eq.~\ref{eq:tdse} then follows stochastic rules, with a discretized representation of the state at each point in imaginary time. After sufficient time, the aim is to sample from a distribution representing the ground state, corresponding to the FCI wavefunction and corresponding basis set correlation energy. Instead of directly sampling the exponential propagator given in Eq.~\ref{eq:propagate}, FCIQMC stochastically simulates the action of a linearized propagator which gives the same long-time solution, given by
\begin{equation}
\ket{\Psi_0} \propto \lim_{n\rightarrow \infty} (1-\Delta \tau \hat{H})^n \ket{\Psi(0)},
\label{eq:LinProp}
\end{equation}
where $\Delta\tau$ represents a small timestep, such that $\tau \sim n\Delta\tau$. While this form can be considered a first-order Taylor expansion of the exponential propagator, which will result in different distributions at intermediate time, the long-time distributions will be identical since the dominant eigenvectors of the propagator coincide for a finite basis set.
We now expand the wavefunction parameterization at any time as a linear combination of the complete set of orthogonal $N$-electron Slater determinants in the basis (the FCI expansion), with each determinant denoted by $\ket{D_i}$, resulting in
\begin{equation}
\ket{\Psi(\tau)} = \sum_i C_i(\tau) \ket{D_i}.
\label{eq:fciqmc_wf}
\end{equation}
Substituting this form of $\ket{\Psi(\tau)}$ in Eq.~\ref{eq:tdse} we arrive at the coupled master equations which govern the time-evolution of the CI coefficients,
\begin{equation}
\frac{\partial C_i}{\partial \tau} = - \sum_j (H_{ij} - S\delta_{ij}) C_j.
\label{eq:get_c_0}
\end{equation}
Here an `energy shift', $S$, has been introduced to control the population dynamics of walkers which are used to represent the CI coefficient in the FCIQMC simulation, and is adjusted such that the norm of the resulting ground-state wavefunction is constant in the long-time limit. An alternative derivation of the dynamics without considering an imaginary-time propagation, can be rationalized from the minimization of a Lagrangian functional\cite{Schwarz17} as
\begin{equation}
\mathcal{L}[\Psi] = \bra{\Psi} \hat{H} \ket{\Psi} - S(\braket{\Psi}{\Psi}-A)	.	\label{eq:GSLag}
\end{equation}
The shift parameter, $S$, now takes the form of a Lagrange multiplier, controlling the normalization of the wavefunction to the constant $A$. Taking a steepest-descent approach to the minimization with respect to the wavefunction amplitudes leads to a finite-difference approximation to Eq.~\ref{eq:get_c_0}, as
\begin{equation}
\Delta C_i = -\Delta \tau \sum_j (H_{ij}-S\delta_{ij}) C_j	,	\label{eq:GSmin}
\end{equation}
which defines the forward-iteration evolution of the amplitudes, and is entirely equivalent to the propagator form of Eq.~\ref{eq:LinProp}.

The wavefunction is now discretized into an ensemble of $N_w$ signed walkers, each of which is assigned to a particular Slater determinant. In the long-time, large-walker number limit of the simulated dynamics, the distribution of walkers will be proportional to the FCI coefficients in Eq.~\ref{eq:fciqmc_wf}.
The population of walkers evolves in every time interval, $\Delta \tau$, according to the propagation of Eq.~\ref{eq:GSmin}, following three stochastic steps:
\begin{enumerate}
\item {\em {Spawning step:}} Each walker selects a new determinant, $\ket{D_j}$, connected to its current determinant, $\ket{D_i}$, by a single application of $\hat{H}$.
This selection is made with a normalized probability $p_{gen}(j|i)$.
Following this selection, a new walker is spawned on $\ket{D_j}$ with a signed probability of $p_s=-\Delta\tau H_{ij}/p_{gen}(j|i)$.
\item {\em {Death step:}} Each occupied determinant, $\ket{D_i}$, attempts to modify its amplitude with a probability $p_d=\Delta\tau (H_{ii}-S)C_i$.
\item {\em {Annihilation step: }} In the final step, newly-created walkers are combined with existing ones, either adding up or cancelling out, depending on their respective signs. 
\end{enumerate}
The energy of the system can be obtained from a projected estimate,
\begin{equation}
E(\tau) = \frac{\bra{\Psi_T}\hat{H}\ket{\Psi(\tau)}}{\braket{\Psi_T}{\Psi(\tau)}},
\end{equation}
where $\ket{\Psi_T}$ is a deterministic trial wave function with a non-zero overlap with $\ket{\Psi_0}$. This will give the exact energy when $\ket{\Psi(\tau)}$ represents an eigenstate of the system\cite{Petruzielo2012}.
However, the energy can alternatively be calculated from the reduced density matrices (RDMs) as a trace with the Hamiltonian matrix\cite{Overy2014,Overy2014t}, corresponding to a pure (symmetric), variational estimate of the ground-state energy.
The sampling of unbiased RDMs via two independent `replica' simulations running in parallel is discussed in more detail in section~\ref{ssec:trdms}.


Further advancements in the FCIQMC algorithm have dramatically improved the efficiency of the method compared to the native algorithm described above. The two main developments in this regard is the addition of the `initiator' approximation\cite{Cleland2010,Cleland2011}, and the semi-stochastic adaptation\cite{Petruzielo2012,Overy2014,Blunt2015c}. The initiator approximation adapts the spawning criteria with the additional requirement for a successful spawning event to an unoccupied determinant conditional on $|N_i|\geqslant n_a$, where $|N_i|$ is the total number of walkers on the parent determinant, and $n_a$ is the initiator parameter, generally set to 3. This will bias the simulation for low walker populations, but results in convergence to the exact result in the limit of increasing population. In the semi-stochastic adaptation, a small number of significant determinants are chosen, and their amplitudes propagated exactly according to Eq.~\ref{eq:LinProp}. This substantially reduces the random errors associated with the dynamic, and reduces the time required to converge the result to a given accuracy. This also allows for a finer resolution of probability amplitudes, with real amplitudes used until they are too low, at which point a stochastic rounding of the total determinant weight occurs to maintain a sparse representation of the state. Without exception, these adaptations are performed for all calculations in this work, and entirely equivalent changes are made for the response dynamic.

\subsection{Reduced Density Matrices within FCIQMC}
\label{ssec:trdms}

In this section, we describe the stochastic accumulation of the RDMs in more detail, since it represents a key step in the sampling of the response properties. These have previously been used for the computation of molecular properties such as (transition) dipole moments\cite{Blunt2017b}, analytic nuclear forces\cite{Thomas2015} and explicitly correlated corrections to the wavefunction\cite{Booth2012a}, in addition to calculating a variational energy estimate within FCIQMC.
A general two-body reduced density matrix is defined as
\begin{equation}
\Gamma^{mn}_{pq,rs} = \bra{\Psi^m}a^\dagger_p a^\dagger_q a_s a_r \ket{\Psi^n},
\end{equation}
where \textit{p}, \textit{q}, \textit{r} and \textit{s} index spin-orbital degrees of freedom. The definition here also allows for the construction of the (transition) RDM between two potentially different states, \textit{m} and \textit{n}.
The one body reduced density matrix is connected to its two body counterparts through its definition as
\begin{eqnarray}
\gamma^{mn}_{p,q}  &=&  \bra{\Psi^m}a^\dagger_p a_q \ket{\Psi^n} \nonumber \\ 
&=& \frac{1}{N-1} \sum_a \Gamma^{mn}_{pa,qa}.
\end{eqnarray}

The FCIQMC coefficients $C^n_i$, are used to obtain the reduced density matrix as
\begin{equation}
\Gamma^{mn}_{pq,rs} = \sum_{ij} E[C_i^m] E[C_j^n] \bra{D_i}a^\dagger_p a^\dagger_q a_s a_r \ket{D_j} ,
\label{eq:est_biased}
\end{equation}
where $E[\dots]$ denotes an expectation value, which can be estimated by an average over the simulation, after convergence has been reached.
The difficulty in a naive sampling of Eq.~\ref{eq:est_biased} is that access to the elements of $E[C^n_i]$ is not possible without an undesirable histogramming of the sampled wavefunction over time. Furthermore, it is not possible to average the samples of the full RDM without introducing a systematic bias at finite walker numbers, since the average of the product $E[ C^m_i \times C^n_j]$ will have biased mean, due to the fact that correlations exist between the sampling of the two terms at each point in time.


This problem is solved with replica sampling in the FCIQMC calculation\cite{Overy2014}. 
With replica sampling, an additional, independent sampling of the same state is performed concurrently. 
As these two simulations for all desired states are performed independently, the random variables $C^m_i$ and $C^n_j$ will be uncorrelated, with zero \mbox{(co-)variance} between the sampled amplitudes. Since expectation values are always quadratic functions of the amplitudes, only two replicas of each state are required, denoted by $C^{1,m}_i$  and  $C^{2,m}_i$ for the amplitude on configuration $i$, state $m$, and replica $1$ and $2$ respectively. This allows for two statistically independent and unbiased estimates of each RDM to be expressed in a bilinear form\cite{Zhang93}, without bias, as
\begin{equation}
\Gamma^{mn}_{pq,rs} = E \bigg[  \sum_{ij} C_i^{1,m} C_j^{2,n} \bra{D_i}a^\dagger_p a^\dagger_q a_s a_r \ket{D_j}   \bigg],	\label{eq:rdm1}
\end{equation}
and
\begin{equation}
\Gamma^{mn}_{pq,rs} = E \bigg[  \sum_{ij} C_i^{2,m} C_j^{1,n} \bra{D_i}a^\dagger_p a^\dagger_q a_s a_r \ket{D_j}   \bigg].	\label{eq:rdm2}
\end{equation}
Expectation values can be averaged from these two estimates due to their independence, resulting in a reduction of the resultant stochastic errors, offsetting the additional cost of propagation of this second walker distribution.


The definitions of the general RDMs in Eqs.~\ref{eq:rdm1} and \ref{eq:rdm2} will result in an {\emph{unnormalized}} density matrix, due to the fact that the FCIQMC wavefunction itself is not normalized.
Whilst normalization of symmetric RDMs can be performed by enforcing known trace relations of the matrices, this is more difficult for transition RDMs, where the trace of each density matrix should be zero. Instead, using the definition of normalized FCIQMC wavefunction as $\frac{1}{A^{R,m}} E[\ket{\Psi^{R,m}}]$ for replica $R$ and state $m$, the normalization of the general 2-body RDM is made with the following scheme\cite{Blunt2017b},
\begin{equation}
\Gamma^{mn}_{pq,rs} [1] = \frac{A^{1,m}A^{2,n}}{\sqrt{A^{1,m}A^{2,m}A^{1,n}A^{2,n}}} \Gamma^{mn}_{pq,rs}	\label{eq:rdmnorm1}
\end{equation}
\begin{equation}
\Gamma^{mn}_{pq,rs} [2] = \frac{A^{2,m}A^{1,n}}{\sqrt{A^{1,m}A^{2,m}A^{1,n}A^{2,n}}} \Gamma^{mn}_{pq,rs},	\label{eq:rdmnorm2}
\end{equation}
which is correct under the assumption that the normalization of the two replicas for each state is equal.

The sampling of Eqs.~\ref{eq:rdm1} and \ref{eq:rdm2} each iteration appears as a double sum over all walkers. However, this can be avoided by a statistical sample of the estimates each iteration. During the course of each spawning event, determinants connected via single and double excitations are generated, and these are used as a sample of the full set of terms required in the RDM sampling, after an appropriate unbiasing for the possibility of their generation. It should be noted that the statistical sampling of the RDMs is only included between {\emph{successful}} spawning events, and as such, it is important that all double excitations that need to be included in the sampling of the RDMs are connected via a non-zero matrix element of the operator which they are sampled from (in the case of the ground-state, this is the Hamiltonian). To minimize this impact, it is ensured that all terms within the deterministic space of the semi-stochastic adaptation are also included, which is designed to include obvious edge cases such as the effect of Brillouins theorem on ensuring that single excitations of the Hartree--Fock determinant are zero. 
In addition to this stochastic sampling of the off-diagonal determinant connections, the diagonal part of the RDMs are sampled exactly, each time a determinant becomes unoccupied.
If these cases are taken into account, it has been shown that the statistical sampling of terms is practically unbiased, and the algorithm therefore allows for the RDM estimators to be computed with relatively little overhead or penalty to the parallelism of the method\cite{Overy2014,Blunt2017b}. 
\section{Theory}
\label{sec:theory}
\subsection{Linear response formalism}
\label{ssec:lrf}

The formalism of response theory allows for the analytic solution to changes in expectation values of a system in terms of a perturbative expansion in the applied perturbation, ${\hat{V}}$.
We briefly review the formalism here mostly in the context of a static perturbation, though more thorough and general expositions are available in Refs.~\onlinecite{Monkhorst1977,Christiansen:1998p36,Helgaker2012}.
Under the assumption of an arbitrary static perturbation which couples linearly to the zeroth-order Hamiltonian with strength $\lambda$, we can write the Hamiltonian as

\begin{equation}
\hat{H}(\lambda)=\hat{H}_0 + \lambda \hat{V}	.
\end{equation}
The energy of this system can be written as 
\begin{equation}
E(\lambda) = \langle \Psi(\lambda) | \hat{H}_0 + \lambda \hat{V} | \Psi(\lambda) \rangle .	\label{eq:pert_e}
\end{equation}
As a specific example, a perturbation can be considered as the application of a static electric field, which under the dipole approximation couples linearly with the strength of the field, ($\lambda=\epsilon$), with the perturbation being the dipole moment operator (${\hat V}={\hat \mu}$). The dipole moment can then be decomposed into a permanent dipole ($\mu^{(0)}_x$), and field-induced dipole contributions, each corresponding to a higher-order response of the system to the field strength $\epsilon$, as
\begin{equation}
\mu_x = \mu^{(0)}_x + \sum_y \alpha_{xy} \epsilon_y + \dots,
\end{equation}
where $\alpha_{xy}$ is the static dipole polarizability of the system. With a Taylor expansion argument, each moment can also be equated to a derivative of the energy of the system with respect to the applied field, as
\begin{equation}
\mu^{(0)}_x = \left. -\frac{\partial E}{\partial \epsilon_x} \right|_{\forall\epsilon={\bf{0}}},	\label{eq:dipole}
\end{equation}
\begin{equation}
\alpha_{xy} = \left. -\frac{\partial^2 E}{\partial \epsilon_x \partial \epsilon_y} \right|_{\forall\epsilon={\bf{0}}}.	\label{eq:pol}
\end{equation}
Analogous perturbative expansions can be set up for other applied fields or expansions of other perturbative contributions to the Hamiltonian. Furthermore, the time-dependent case can be constructed analogously, whereby a monochromatic time-dependent perturbation of frequency $\omega$ can be written as
\begin{equation}
\hat{V}(t) = {\hat V}(e^{-i\omega t}+e^{+i\omega t}).
\end{equation}
This allows for the construction of a quasi-energy eigenvalue equation, which can again be expanded in powers, and which reduces to the static limit as $\omega \rightarrow 0$. We will leave the construction of dynamic response functions for future work, while we will focus in this work on the static case.

In order to compute the response quantities allowing for generalization of Eqs.~\ref{eq:dipole} and \ref{eq:pol}, we can differentiate Eq.~\ref{eq:pert_e}, which (for real wavefunctions) gives
\begin{equation}
\frac{d E}{d \lambda} = 2\left\langle \frac{\partial \Psi}{\partial \lambda} \right| (\hat{H}_0 +\lambda  \hat{V}) | \Psi \rangle + \langle \Psi | \hat{V} | \Psi \rangle	\label{eq:energyderiv}
\end{equation}
and taking the perturbation strength to zero gives
\begin{equation} 
\left. \frac{d E}{d \lambda} \right|_{\lambda=0} = 2 \left\langle \frac{\partial \Psi}{\partial \lambda} \right| {\hat H}_0 | \Psi_0 \rangle	+ \langle \Psi_0 | \hat{V} | \Psi_0 \rangle ,	\label{eq:1storderres}
\end{equation}
where $\Psi_0$ is the solution to the unperturbed problem.
We can write derivatives of the wavefunction with respect to the perturbation strength as
\begin{equation}
\frac{\partial \Psi}{\partial \lambda} = \sum_i \frac{\partial \Psi}{\partial C_i} \frac{\partial C_i}{\partial \lambda}
\end{equation}
and
\begin{equation}
\frac{\partial^2 \Psi}{\partial \lambda^2} = \sum_{ij} \frac{\partial^2 \Psi}{\partial C_j \partial C_i} \frac{\partial C_j}{\partial \lambda}\frac{\partial C_i}{\partial \lambda} + \sum_i \frac{\partial^2 C_i}{\partial \lambda^2}\frac{\partial \Psi}{\partial C_i},	\label{eq:sec_deriv_psi}
\end{equation}
where $C_i$ denotes the wavefunction parameters associated with $\Psi$, which in the case of the FCIQMC parameterization are given in Eq.~\ref{eq:fciqmc_wf}. Collecting these parameters in a vector ${\bf C}$ for compactness, we can then write Eq.~\ref{eq:1storderres} as 
\begin{equation}
\left. \frac{d E}{d \lambda} \right|_{\lambda=0} = 2\frac{\partial \bf{C}}{\partial \lambda} \left\langle \frac{\partial \Psi}{\partial {\bf C}} \right| {\hat H}_0 | \Psi_0 \rangle + \langle \Psi_0 | \hat{V} | \Psi_0 \rangle	.	\label{eq:firstordereq}
\end{equation}
Since the FCIQMC approach aims to minimize the unperturbed energy with respect to all coefficients, at least in an averaged sense, we can use the stationarity of the unperturbed wavefunction to assert that
\begin{eqnarray}
\left. \frac{\partial E}{\partial {\bf C}}\right|_{\lambda=0} &=& 2 \left. \left\langle \frac{\partial \Psi}{\partial {\bf C}} \right| {\hat H}_0 + \lambda \hat{V} | \Psi \rangle \right|_{\lambda=0} \label{eq:Firstorderorth1} \\
&=& 2\left\langle \frac{\partial \Psi}{\partial {\bf C}} \right| {\hat H}_0 | \Psi_0 \rangle = 0	.	\label{eq:Firstorderorth}
\end{eqnarray}
Using Eqs.~\ref{eq:Firstorderorth} and \ref{eq:firstordereq} gives the familiar expression for the first-order response of the energy to a perturbation as
\begin{equation}
\left. \frac{d E}{d \lambda} \right|_{\lambda=0} = \langle \Psi_0 | {\hat V} | \Psi_0 \rangle	,
\end{equation}
which for the example of the applied electric field, gives the permanent dipole moment in Eq.~\ref{eq:dipole}.

For the second-derivative properties, we can differentiate Eq.~\ref{eq:energyderiv} once more, to give
\begin{eqnarray}
\frac{d^2E}{d \lambda^2} =& 2\left\langle \frac{\partial^2 \Psi}{\partial \lambda^2}\right| {\hat H}_0 + \lambda{\hat V} | \Psi\rangle + 2\left\langle\frac{\partial \Psi}{\partial \lambda} \right| {\hat H} + \lambda {\hat V} \left| \frac{\partial \Psi}{\partial \lambda} \right\rangle	 	\nonumber \\
&+ 2\left\langle\frac{\partial \Psi}{\partial \lambda}\right| {\hat V} | \Psi \rangle  ,
\end{eqnarray}
which in the limit of vanishing perturbation strength, gives
\begin{eqnarray}
\left.\frac{d^2E}{d \lambda^2}\right|_{\lambda=0} =& 2\left\langle\frac{\partial^2\Psi}{\partial \lambda^2} \right| {\hat H}_0 | \Psi_0\rangle +   2\left\langle \frac{\partial \Psi}{\partial \lambda} \right| {\hat H}_0 \left| \frac{\partial \Psi}{\partial \lambda}\right\rangle  \nonumber \\
&+  2\left\langle \frac{\partial \Psi}{\partial \lambda} \right| {\hat V} | \Psi_0 \rangle .
\end{eqnarray}
Using Eq.~\ref{eq:sec_deriv_psi}, the above equation can be written in terms of variation in the wavefunction parameters, giving
\begin{eqnarray}
\left.\frac{d^2E}{d \lambda^2}\right|_{\lambda=0} &= 2\frac{\partial^2 {\bf C}}{\partial \lambda^2} \left\langle\frac{\partial\Psi}{\partial \bf{C}} \right| {\hat H}_0 | \Psi_0\rangle +2\left(\frac{\partial \bf{C}}{\partial \lambda}\right)^2 \left\langle\frac{\partial^2\Psi}{\partial \bf{C}^2} \right| \hat{H}_0 | \Psi_0 \rangle \nonumber \\
& + 2\left\langle \frac{\partial \Psi}{\partial \lambda} \right| {\hat H}_0 \left| \frac{\partial \Psi}{\partial \lambda}\right\rangle + 4\left\langle \frac{\partial \Psi}{\partial \lambda} \right| {\hat V} | \Psi_0 \rangle.	\label{eq:secorderexpand}
\end{eqnarray}
This can now be simplified as in the first-order response case, by returning to the stationarity conditions of Eqs.~\ref{eq:Firstorderorth1} and \ref{eq:Firstorderorth}. Differentiating Eq.~\ref{eq:Firstorderorth1} with respect to $\lambda$ gives an expression which must also equal zero due to stationarity,
\begin{eqnarray}
&\left.\frac{\partial}{\partial \lambda}\left\langle\frac{\partial \Psi}{\partial {\bf C}} \right| {\hat H}_0 + \lambda \hat{V} | \Psi \rangle \right|_{\lambda=0} = \left(\frac{\partial {\bf C}}{\partial \lambda} \right) \left\langle \frac{\partial^2\Psi}{\partial{\bf C}^2} \right| {\hat H}_0 | \Psi_0 \rangle \nonumber \\
&+ \left\langle\frac{\partial \Psi}{\partial \bf{C}} \right| {\hat V} | \Psi_0 \rangle + \left(\frac{\partial \bf{C}}{\partial \lambda} \right) \left\langle\frac{\partial \Psi}{\partial {\bf C}} \right| {\hat H}_0 \left| \frac{\partial \Psi}{\partial {\bf C}} \right \rangle = 0.	\label{eq:SecOrderorth}
\end{eqnarray}
Combining Eqs.~\ref{eq:Firstorderorth} and \ref{eq:SecOrderorth} into Eq.~\ref{eq:secorderexpand}, results in the expression for second-order properties as
\begin{equation}
\left.\frac{d^2E}{d \lambda^2}\right|_{\lambda=0} = 2\left\langle\frac{\partial\Psi}{\partial\lambda}\right|{\hat V}|\Psi_0\rangle	.	\label{eq:secorderprop}
\end{equation}
For the example of the polarizability given in Eq.~\ref{eq:pol}, this then results in
\begin{equation}
\alpha_{ij} = -2\left\langle\frac{\partial\Psi}{\partial \epsilon_i} \right| \hat{\mu}_j | \Psi_0 \rangle	. \label{eq:pol2}
\end{equation}

To solve these equations requires the explicit computation of the response vector, $\frac{\partial\Psi}{\partial\lambda}$. To achieve this, we can return to Eq.~\ref{eq:SecOrderorth}, written as
\begin{eqnarray}
&\left(\left\langle\frac{\partial^2\Psi}{\partial\bf{C}^2}\right|{\hat H}_0|\Psi_0\rangle + \left\langle\frac{\partial\Psi}{\partial\bf{C}}\right| {\hat H}_0 \left|\frac{\partial\Psi}{\partial{\bf C}}\right\rangle\right)\left(\frac{\partial {\bf C}}{\partial \lambda}\right) \nonumber \\
&=-\left\langle\frac{\partial\Psi}{\partial{\bf C}} \right| {\hat V} | \Psi_0 \rangle	.	\label{eq:LinEqs}
\end{eqnarray}
Eq.~\ref{eq:LinEqs} represents a set of linear equations to solve for the response vector, and its stochastic numerical solution for a sparse representation of the response vector is the central objective of this work. However, it should be stressed that this is a more general problem, also applicable to higher-order response. For higher order response, it is often necessary to solve for the higher-order response vector. Following a similar derivation as the one outlined for the second-order response, an analogous linear system can be formulated for the response vector $\frac{\partial^2 {\bf C}}{\partial \lambda^2}$, where the right-hand side of the linear system now depends on $-\left\langle\frac{\partial \Psi_0}{\partial {\bf C}} \right| {\hat V} \left| \frac{\partial \Psi_0}{\partial \lambda}\right\rangle$, which can be obtained from the solution to the first-order response vector equation. This sets up a hierarchy of linear response equations, which can all be solved simultaneously within the framework described below. Wigner's `$(2n+1)$' rule stipulates that in order to calculate the energy response up to order $2n+1$, the response vector up to order $n$ is required\cite{Helgaker1989,Christiansen:1998p36}. In this work, we will not numerically demonstrate beyond the first-order response vector (required for up to third-order energy derivatives), but leave this investigation for future work.

\subsection{Sampling Response Properties in FCIQMC}
\label{ssec:lre}

We now turn to the numerical stochastic solution to the response equations within the framework of FCIQMC. We consider the response of the FCIQMC state in the large walker limit, meaning that any residual initiator error in the description of the zeroth order state is neglected. This means that the response may differ from the finite field value for small walker numbers, but become increasingly accurate to exactness as the number of walkers increases. Due to the difficulty in formulating the initiator approximation as a strict constraint on the wave function ansatz, it is easier, and correct in the large walker limit, to simply formulate the response in the absence of initiator error, allowing us to follow on directly from Eq.~\ref{eq:LinEqs}. As this requires the solution to a linear equation, rather than an extremal eigenvalue/vector pair of a matrix as required for the ground-state algorithm, changes are required of the stochastic rules governing the walker propagation. However, similar modifications to stochastic algorithms have been performed before, in the context of Multi-State Quantum Monte Carlo\cite{Tenno13}, and a perturbative coupling within FCIQMC\cite{Jeanmairet17}. In order to define the response vector within FCIQMC, we start by writing the corresponding wavefunction in presence of the perturbation as
\begin{eqnarray}
\ket{\Psi} &=& \ket{\Psi^{(0)}} + \lambda \ket{\Psi^{(1)}} + \frac{1}{2} \lambda^2 \ket{\Psi^{(2)}} + \cdots \nonumber \\
&=& \sum_i C_i \ket{D_i},
\end{eqnarray}
where the linear amplitudes can also be expanded in perturbative orders as
\begin{equation}
C_i = C^{(0)}_i  + \lambda C^{(1)}_i + \frac{1}{2} \lambda^2 C^{(2)}_i + \cdots.
\end{equation}
With this definition, the first-order response of the wavefunction  can now be written as
\begin{equation}
\ket{\Psi^{(1)}} = \left.\frac{\partial \Psi}{\partial \lambda} \right|_{\lambda=0} =  \sum_i \frac{\partial C_i}{\partial \lambda} \frac{\partial \Psi}{\partial C_i} = \sum_i C_i \ket{D_i}
\label{eq:ResWave}
\end{equation}
Substituting this linear parameterization for the wavefunction forms of both the zeroth-order and the response state into Eq.~\ref{eq:LinEqs} results in
\begin{equation}
(\hat{H}_0-E_0) |\Psi^{(1)} \rangle = -{\hat{Q}}\hat{V}\ket{\Psi^{(0)}}	,	\label{eq:ResMaster}
\end{equation}
where the first term of Eq.~\ref{eq:LinEqs} is zero due to the linearity of the wavefunction. Two additional terms have been introduced to enforce the desired intermediate normalization of the response functions. $\hat{Q}$ is a projection operator, defined as $\hat{Q}=1-\ket{\Psi^{(0)}}\bra{\Psi^{(0)}}$, ensuring that the response vector is orthogonal to the zeroth-order wavefunction by projecting it out from the response. Furthermore, the $E_0$ contribution of Eq.~\ref{eq:ResMaster} is also included to ensure that the expression in Eq.~\ref{eq:pert_e} is appropriately normalized. In order to derive an iterative scheme to solve these equations, we define a modified Lagrangian, which will result in the solution to Eq.~\ref{eq:ResMaster} as its minimum, analogous to the approach taken for Eq.~\ref{eq:GSLag}. This can be achieved with the form
\begin{equation}
\mathcal{L}[\Psi^{(1)};\Psi^{(0)}] = \frac{1}{2}\bra{\Psi^{(1)}} \hat{H}_0-S \ket{\Psi^{(1)}} - \bra{\Psi^{(1)}} \hat{V} \ket{\Psi^{(0)}}	,	\label{eq:ResLag}
\end{equation}
where $S$ is the same energy shift used to define the zeroth-order energy. It is simple to show that the (unique) global minimum of this equation will give the $\ket{\Psi^{(1)}}$ defined in Eq.~\ref{eq:ResMaster}. In order to define an iterative scheme, we again apply a steepest-descent approach to minimize Eq.~\ref{eq:ResLag} with respect to the amplitudes of Eq.~\ref{eq:ResWave}, as
\begin{eqnarray}
\Delta|\Psi^{(1)}\rangle &=& -\Delta\tau\nabla\mathcal{L}[\Psi^{(1)};\Psi^{(0)}]	\\
&=&-\Delta\tau(\hat{H}-S)|\Psi^{(1)}\rangle - \alpha\hat{V}\ket{\Psi^{(0)}}	,	\label{eq:alpha}
\end{eqnarray}
where $\alpha$ is a scalar term introduced to allow for control over the normalization of the sampled response vector, which will be discussed later.
Note that this equation is valid for any choice of $\hat{V}$. The final forward-iteration scheme in component form can then be written as
\begin{equation}
\Delta C^{(1)}_i =  \underbrace{-\Delta\tau\sum_j(H_{ij}-S\delta_{ij})C^{(1)}_j}_{`Hamiltonian' dynamics} - \underbrace{\Delta\tau \alpha V_{ij}C^{(0)}_j}_{`Perturbation' dynamics}	.	\label{eq:ResMaster2}
\end{equation}

In order to construct the stochastic algorithm, it is necessary to discretize both the ${\bf C}^{(0)}$ and ${\bf C}^{(1)}$ amplitudes into a sparse walker representation. It should be noted that the dynamics have only a linear dependence on each walker distribution, so that the modification of the wavefunction amplitudes into random variables will not introduce bias in their solution. In defining these discrete, signed walkers, we can simulate the convergence of the dynamics of the zeroth-order state according to Eq.~\ref{eq:GSmin} and the response state according to Eq.~\ref{eq:ResMaster2} at the same time. While the dynamics of the zeroth-order state walker distribution are entirely independent from the response state dynamics, an interaction between the two walker ensembles takes place via the second term of Eq.~\ref{eq:ResMaster2}, which modifies the dynamics of the response state walkers. The rules for the response state walker dynamics are therefore as follows:
\begin{enumerate}
\item {\em {`Hamiltonian' spawning step:}} For each response-state walker on each occupied determinant, $\ket{D_i}$, a spawning step is made equivalently to the dynamics of the zeroth-order state, to a randomly chosen determinant $\ket{D_j}$, updating the response-state walker distribution.
\item {\em {`Perturbation' spawning step:}} For each walker in the {\em zeroth-order} walker ensemble residing on an occupied determinant, $\ket{D_i}$, a random determinant is chosen, $\ket{D_k}$, which is connected via the rank and symmetry of the perturbation operator ${\hat{V}}$, with a normalized generation probability $p^V_{gen}(k|i)$. A walker is semi-stochastically generated in the {\em response-state} walker ensemble with a signed probability of $-\Delta\tau \alpha V_{ik}/p^V_{gen}(k|i)$.
\item {\em {`Hamiltonian' death step:}} Equivalently to the zeroth-order dynamics, each occupied response determinant, residing on $\ket{D_i}$, semi-stochastically alters its cumulative signed weight according to a probability $p_d=\Delta\tau (H_{ii}-S)C_i^{(1)}$.
\item {\em {`Perturbation' death step:}} On each occupied {\em zeroth-order state} determinant, response-state walkers are semi-stochastically created on $\ket{D_i}$ with a signed probability of $-\Delta\tau \alpha V_{ii}C^{(0)}_i$.
\item {\em {Annihilation step:}} Separately, the walkers on each state are collected and cancelled, depending on their respective signs.
\end{enumerate}
The algorithm for the above steps is more efficient if the occupied determinants of both states are stored together, with more details of optimal datastructures found in Ref.~\onlinecite{Booth2014}. In practice, this is achieved with the hashing algorithm, and then running through the occupied determinants of both states is simple, and allows for efficient annihilation and merging of the walker lists. Furthermore, it should be noted that following the semi-stochastic algorithm detailed in Ref.~\onlinecite{Blunt2015c}, we allow for the exact propagation of these rules within a chosen `deterministic' subspace of the determinant space to reduce stochastic errors. In this work, this space is chosen to simply be the single and double excitation space of the Hartree--Fock determinant.

It should be noted that the above algorithm and equations are missing one step, which is that the projection operator, ${\hat{Q}}$ is missing from Eq.~\ref{eq:ResMaster2}, and the algorithm above. This is exactly remedied, via a semi-stochastic application of the Gram-Schmidt algorithm, which has previously been successfully employed to ensure orthogonality between states for excited eigenstate calculations within FCIQMC\cite{Blunt2015a}. Even though the zeroth-order state is only known stochastically at any one iteration, this projection operator can still be constructed and applied in a practically unbiased fashion. This necessitates a final step each iteration:
\begin{enumerate}
\setcounter{enumi}{5}
\item {\em {Orthogonalization step:}} By running over the occupied states for both walker distributions, a Gram-Schmidt step can be used to semi-stochastically orthogonalize the response state walker distribution with respect to the zeroth-order state walker distribution. This leaves the zeroth-order state walkers unchanged.
\end{enumerate}
The response due to perturbation operators which are not totally symmetric are also often desired, and these perturbations may even break spin or particle number symmetry. In these cases, the perturbation will connect the zeroth-order state symmetry sector to another sector, and therefore the response walkers will live in an entirely different symmetry sector to the zeroth-order state walkers. In this case, both the orthogonalization step and the perturbation death step are unnecessary, as the states will always be orthogonal by symmetry, and the death step will also be zero, as there will be no diagonal part to the perturbation operator.

The perturbation spawning steps require an excitation generation algorithm which corresponds to the random excitation between determinants, constrained by the rank and symmetry of the perturbation applied. If the perturbation is totally symmetric and no more than two particle, it is possible to simply reuse the excitation generators used for the Hamiltonian sampling\cite{Booth2014}. However, for many perturbations, it is possible to have a significantly more efficient algorithm. For the one-body perturbation operators used in this work, we can devise an `exact' random perturbation excitation generation algorithm. In this, excitations are randomly generated with a probability exactly given by the modulus of the perturbation matrix element between the two excitations. This is simply obtained by precomputing a normalized cumulative distribution function for each orbital, as given by the modulus of the perturbation operator. This is then used for the picking of the electrons and holes in a given determinant to excite from/to respectively. This maximises the efficiency of the sampling of the perturbation spawning step, with no additional cost during the run and minimal overhead in initialization.

The response simulation is initialized without any walkers, and the zeroth-order state walker distribution initialized as normal (generally with a single walker placed at the reference determinant). This allows the response state occupation to arise due to the `perturbation' spawning/death steps. Unlike the zeroth-order state dynamics where the normalization of the walkers is controlled by the shifting of the Hamiltonian operator, no analogous shifting can be used for the normalization of the response walkers, since its normalization is fixed by the choice of zeroth-order normalization. This is important to manage computational resources, and control the effort expended in sampling the response state wavefunction distribution. We achieve this by scaling the magnitude of the perturbation operator with the $\alpha$ parameter introduced in Eq.~\ref{eq:alpha}. This will scale the normalization of the response state, but also the response expectation values, and therefore needs to be taken into account in the evaluation of all response properties. We allow for the number of response walkers ($N_w^{(1)}$) to be some multiple of the number of zeroth-order state walkers, such that
\begin{equation}
\frac{N_w^{(1)}}{N_w^{(0)}}=f^{(1)}	.
\end{equation}
This is achieved by every update cycle (25 or so iterations), updating $\alpha$ as
\begin{equation}
\alpha=\frac{N_w^{(0)} f^{(1)}}{N_w^{(1)}}	,
\end{equation}
and then fixing it when walker numbers have stabilized and statistics are being accumulated.
It would be possible (and probably reasonable) to choose a different number of walkers to sample the zeroth-order state and response state distribution, however in this initial work, we choose $f^{(1)}=1$, ensuring that we have the same numbers of walkers sampling the two state distributions.

Unbiased higher-order response vectors can also be simulated in the algorithm detailed above, as well as the response of a stochastically sampled excited state, where the zeroth-order state is not the ground state walker distribution. The ability to sample excited states is detailed in Ref.~\onlinecite{Blunt2015a}, where multiple walker distributions are set up and orthogonalized against each other to allow for a pure-state sampling of low-lying eigenstates. An additional walker distribution can be used to sample the response of any of these desired states, by ensuring that the zeroth-order state in the above equations denote instead the excited state walker distribution to which the perturbation is applied to. Furthermore, a stochastic hierarchy of higher-order response states can be set up by introducing additional walker distributions. To sample the $n^{\textrm{th}}$-order response vector will involve the perturbation in Eq.~\ref{eq:ResMaster2} no longer acting on the zeroth-order state, but instead on the $(n-1)^{\textrm{th}}$-order response walkers. It should be noted in this case that the $S$ value and orthogonalization should still be applied with respect to the zeroth-order state. Investigations of these higher-order response quantities will be reserved for future work.

Finally, it is important to be able to compute symmetric and asymmetric expectation values between the response and pure state walker distributions, as required to evaluate e.g. Eq.~\ref{eq:pol2}. These are computed via the accumulation and long-time average of the reduced density matrices and/or transition reduced density matrices (tRDMs) between the different states, as detailed analogously for eigenstate walker distributions in Ref.~\onlinecite{Blunt2017b} and briefly reviewed in section~\ref{ssec:trdms}. 
In this, the sampling of the RDMs occurs `through' an operator. By this, we mean that the choice of determinants to include in the sampling of Eqs.~\ref{eq:rdm1} and \ref{eq:rdm2} are given by the `spawning' steps of the walker dynamic algorithm, in order to efficiently compute these quantities. 

In the response sampling, we now have additional spawning steps, which can be used to increase the number of samples of each desired RDM, by including sampling of states connected via the ${\hat{V}}$ in the `perturbation' spawning. This is essential for perturbations which are not totally symmetric, where the transition RDM required connects states of different symmetry, and therefore the Hamiltonian spawning will never result in contributions between the two states. For the results in this paper, for totally symmetric perturbations, we sample the required tRDMs only through the Hamiltonian sampling, while for non-symmetric perturbations, these are sampled entirely through the perturbation sampling. In the future, efficiency gains are likely for symmetric perturbations by sampling the tRDMs through all possible stochastic spawning steps with the same symmetry as the perturbation.

As with the ground-state (t)RDMs, there is the potential for a bias to result in the averaging of the RDMs due to their quadratic dependence on the random variables and neglect of the (co)variances which result if only one walker distribution is used for each state. These are likely to be far reduced for transition RDMs as required for second-order response quantities, since there is no squared single determinant weights required in their computation. However, there are still formally correlations between the walker distributions due to their interaction via the perturbation. It can be ensured that this potential bias is formally eliminated by using two walker replicas for both the ground-state sampling and response states. In section~\ref{sec:results} we will investigate the necessity for two replicas for each state to avoid this potential bias. 

Transition RDMs are then evaluated as described in Eqs.~\ref{eq:rdmnorm1} and \ref{eq:rdmnorm2}. For the example of polarizabilities given in Eq.~\ref{eq:pol2}, response properties can therefore be calculated as
\begin{equation}
\alpha_{xy}=- \frac{1}{2}\sum_{pq} \left[ \hat{x}_{pq} \gamma_{p,q}^y+ \hat{y}_{pq} \gamma_{p,q}^x  \right],	\label{eq:pol_gamma_1}
\end{equation}
with the $\gamma_{p,q}^y$ obtained as
\begin{equation}
\gamma_{pq}^y=\sum_{pq}  \frac{1}{(N-1)} \sum_a \left[ \frac{1}{\alpha_1} \Gamma_{pa,qa}^{(0)(1)}[1]+\frac{1}{\alpha_2}\Gamma_{pa,qa}^{(0)(1)}[2]\right].	\label{eq:pol3}
\end{equation}
Here we have explicitly summed the resulting tRDMs between both the replica pairs (if used) connecting the zeroth-order and response states as defined in Eqs.~\ref{eq:rdm1} and \ref{eq:rdm2}. $\alpha_1$ and $\alpha_2$ correspond to the $\alpha$ values for each respective response replicas, and the response perturbation in the computation of the tRDMs in Eq.~\ref{eq:pol3} is taken to be the ${\hat{y}}$ operator. 

\section{Results and discussion}
\label{sec:results}

The stochastic response algorithm described was implemented within the {\tt NECI} codebase\cite{neci}, where integrals for the electronic Hamiltonian and perturbation were obtained in a Hartree--Fock basis from the {\tt PySCF} program\cite{PySCF}. In this work, we compute the static polarizability tensor of the ground state as an example response property, following the expressions in Eqs.~\ref{eq:pol2} and \ref{eq:pol_gamma_1}. We test the performance of the method on all the $\hat{x}$, $\hat{y}$ and $\hat{z}$ components of the perturbation for a range of heteronuclear diatomic test systems. This will ensure that both symmetric ($\hat{z}$) and non-symmetric ($\hat{x}$ and $\hat{y}$) perturbations are considered within the $C_{2v}$ molecular point group employed, as these require differing sampling procedures for the tRDMs as detailed above. Beyond ensuring the correctness of the algorithm, we aim to focus this initial investigation on three areas:
\begin{enumerate}
\item {\em Convergence with sampling time} 

An important convergence criteria is the number of iterations required to sample the response properties to reach acceptable random error bars. While the uncertainty in Monte Carlo should follow a $\sim\frac{1}{\sqrt{N_{\rm{iter}}}}$ decay of the random errors, this cost is also heavily influenced by the autocorrelation lengths of the simulation, and therefore this convergence will be considered in section~\ref{ssec:rdm-sampling}.
\item {\em Convergence with walker number} 

The number of walkers required to converge desired values is a good proxy for the computational effort of the calculation, with the computational effort scaling linearly with the number of walkers\cite{Booth2014}. In section~\ref{ssec:inc_walkers} we investigate how the different components of the polarizability converge with this number in order to saturate the initiator error in the results and systematically converge to `exact' FCI accuracy.
\item {\em Comparison of sampling schemes} 

In section~\ref{ssec:diff_schemes}, we investigate potential biases arising from different sampling algorithms of the response values. Comparison will be made between two schemes. In the first, the tRDMs are accumulated over the course of the full simulation, and contracted at the end to calculate the response properties. In order to calculate error bars associated with this property, multiple independent simulations are run with different seeds to compute random error bars from entirely independent runs, in a fully uncorrelated way. 

In the second scheme, we contract short-time sampled tRDMs to provide running on-the-fly estimates of the response property directly during the course of a single calculation. While these estimates are serially correlated with each other, a blocking procedure\cite{Flyvbjerg1989} can be run to estimate the true error bar from these correlated estimates. This scheme is advantageous since only a single simulation is required and the convergence of the final property can be monitored during the calculation. We will check that these two schemes are equivalent and do not lead to a bias in the averaged result. Furthermore, we will investigate whether two replicas are required in the algorithm, and determine whether correlations between the zeroth-order state and response walker distributions are small enough to avoid this additional replica cost to eliminate all potential non-linear biases when computing response quantities.

\item {\em Application to CN and NO and comparison to high-level coupled-cluster}

Finally, in section~\ref{ssec:cn_no} we apply the approach to compute the polarizability tensor for the CN and NO radical systems in a d-aug-cc-pVDZ basis set, far beyond those which can be treated by exact FCI approaches. We also compare to high-level coupled-cluster results for the polarizability of these systems, and resolve a source of disagreement between UHF and ROHF-based high-level coupled-cluster linear response approaches for these systems.
\end{enumerate}
\begin{figure*}[]
\begin{center}
\floatbox[{\capbeside\thisfloatsetup{capbesideposition={right,center},capbesidewidth=5cm}}]{figure}[\FBwidth]
{\caption{
Convergence of diagonal components of the polarizability for LiH and BH in aug-cc-pV$X$Z ($X$=D,T,Q) basis sets as the number of sampling iterations is increased at fixed walker number ($10^6$ walkers).  
The polarizabilities are presented as a deviation to the corresponding values after $10^5$ sampling iterations.
}
\label{fig:iter}}
{\includegraphics[trim={0.0cm 0cm 0cm 0cm},clip,scale=0.6]{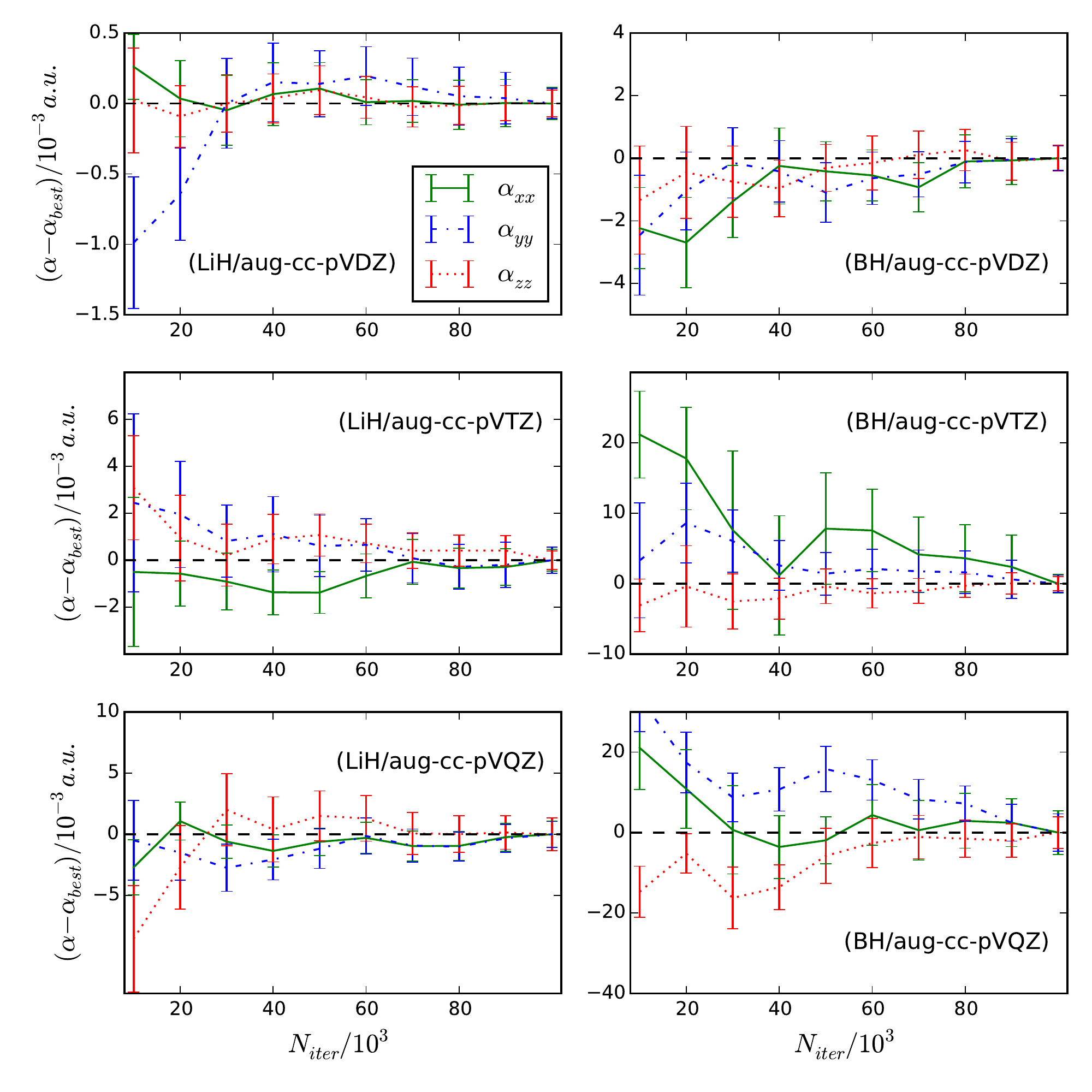}}
\end{center}
\end{figure*}

\begin{figure*}[htbp]
\begin{center}
\floatbox[{\capbeside\thisfloatsetup{capbesideposition={right,center},capbesidewidth=5cm}}]{figure}[\FBwidth]
{\caption{
Convergence of the systematic initiator error for the parallel ($\alpha_{||}$) and perpendicular ($\alpha_{\perp}$) components of the static polarizability for LiH and BH in aug-cc-pV$X$Z ($X$=D,T,Q) basis sets as the number of walkers used in the simulations are increased. 
Results are shifted relative to the value, labelled here as the best value, which correspond to the calculations using the highest number of walkers involved in each of the calculations.
The best results obtained from Coupled Cluster (CC) linear response calculations are also presented. 
For molecules in an aug-cc-pVDZ basis, these are CCSDTQ results, whereas for the rest of the systems these are CCSDT results.
}
\label{fig:walker}}
{\includegraphics[trim={0.0cm 0cm 0.0cm 0cm},clip,scale=0.6]{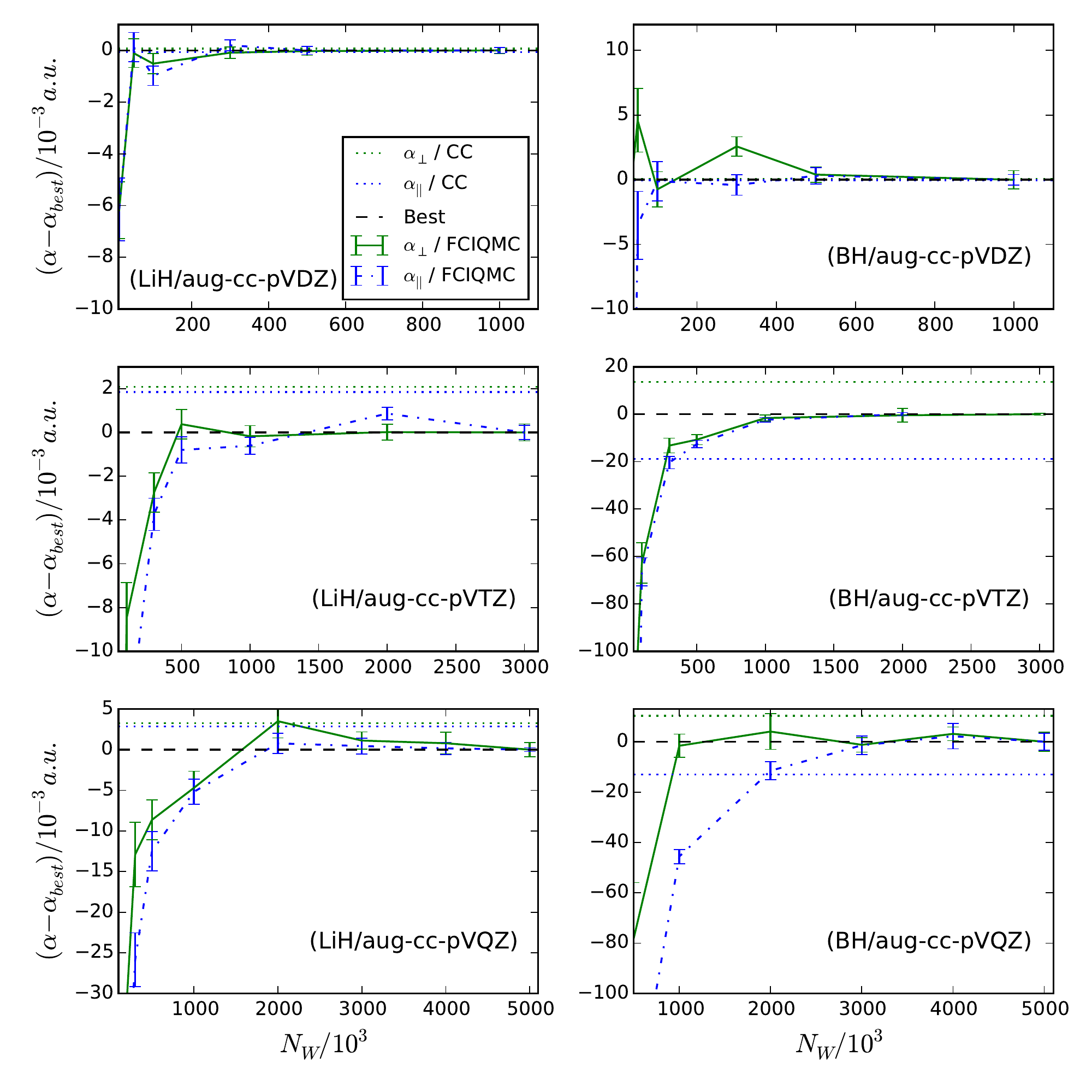}}
\end{center}
\end{figure*}

\begin{table*}[htb]
\begin{center}
\caption{
Final converged estimates for energy, dipole moment and different components of electrical polarizability tensor, calculated for LiH and BH molecules using aug-cc-pV$X$Z ($X$=D,T,Q) basis set. 
The calculations using the aug-cc-pVDZ, aug-cc-pVTZ and aug-cc-pVQZ  basis sets are done with 10$^6$,  $3\times10^6$ and $5\times10^6$ number of walkers, respectively.
 All estimates are obtained after sampling RDMs for  $10^5$ iterations for the aug-cc-pVDZ and aug-cc-pVTZ basis, and for $5\times10^4$ iterations for the aug-cc-pVQZ basis.
Three different schemes are used here to get the energy and property estimators in FCIQMC, details of which are given in the text. 
The best  CC results for aug-cc-pVDZ are given by CCSDTQ, while for the other systems they are obtained from CCSDT linear response values.
}
\label{tab:diff_methods}
\begin{tabular*}{\columnwidth}{@{\extracolsep{\fill}}lcccc} \hline \hline 
Properties & Averaged over multiple runs & Averaged within a single run  & Without replica/Single run &  Best CC results \\ 	\hline  \\
 \multicolumn{5}{c}{LiH/aug-cc-pVDZ} \\ 	[1ex]
 Energy & -8.02122538(3)	 & 	-8.02122539(3) &  -8.02122809(4) & -8.02122537   \\
 $\mu_z$ & 2.328395(2)	 & 2.328394(1) & 	2.328397(2)	&  2.328396 \\
$\alpha_{xx}$	& 30.26148(3)  & 30.2615(1) &	 30.26121(9) 	&	30.261381  \\ 
$\alpha_{yy}$	& 30.26139(7)  & 30.2611(1) &	 30.26146(9) & 30.261381  \\
$\alpha_{zz}$	& 26.10366(5)  & 26.1036(1)	&  26.10354(8)	&	26.103569  \\  [1ex]
 \multicolumn{5}{c}{LiH/aug-cc-pVTZ} \\ 	[1ex]
 Energy & -8.03886622(4)	& 	-8.0388662(1)	& 	-8.0388884(1)	&	 -8.03886596 \\
 $\mu_z$ & 2.309894	(6) &	2.309898(3) &	2.309876(4) &	2.309899 \\
$\alpha_{xx}$	&	30.0059(1) & 30.0057(5) & 30.0047(5) & 30.008131 \\
$\alpha_{yy}$	&	30.0061(2) & 30.0064(3) & 30.0064(6) & 30.008131 \\
$\alpha_{zz}$	&	26.1871(2)	& 26.1871(3) & 26.1872(3)  & 26.188907 \\ [1ex]
 \multicolumn{5}{c}{LiH/aug-cc-pVQZ} \\ 	[1ex]
  Energy & -8.0449978(2) &	-8.0449976(2)	&	 -8.0450286(2)	&		-8.04499753 \\
 $\mu_z$ & 2.305100	(7) & 2.30510(1)	& 2.30495(2) &		2.305101 \\
 $\alpha_{xx}$ & 29.811(3) &	29.807(1)	&	29.8090(9)	 & 29.811040 \\
 $\alpha_{yy}$ & 29.811(1)	&  29.811(1) & 	29.807(2)	 & 29.811040 \\
 $\alpha_{zz}$ & 25.963	(2) &	25.962(1) &	25.9640(6) &  25.963916 \\	
 \multicolumn{5}{c}{BH/aug-cc-pVDZ} \\ 	[1ex]
 Energy & -25.2197069(7) & -25.2197065(6) & -25.219699(1)	&	-25.21970721 \\
 $\mu_z$ &  0.50478(2) & 0.50479(2) & 	0.50469(4) &		0.504772 \\
$\alpha_{xx}$  &	20.2902(2) & 20.2900(8) &	 20.2891(3) & 20.289994 \\
$\alpha_{yy}$  &	20.2905(4) &	20.2899(6) &	 20.2872(3) & 20.289994 \\
$\alpha_{zz}$  &	23.9767(2) &	23.9765(4) &	23.9761(3)  & 23.976487 \\ [1ex]
 \multicolumn{5}{c}{BH/aug-cc-pVTZ}  \\ [1ex]
Energy & -25.243206	(1) & -25.243205(2) &	-25.242261(2)	&	-25.24310758 \\
 $\mu_z$ & 0.52294(2)	&	0.52298(2) &	0.52278(5) &		0.523788 \\
$\alpha_{xx}$	 &	20.7238(8) &	 20.7225(5) &	20.7203(9) &		20.737067 \\ 
$\alpha_{yy}$	 &	20.7251(5) &	 20.7244(4) &	20.7222(9) &		20.737067 \\
$\alpha_{zz}$	 &	23.7294(6) &	 23.7303(4) &	23.720(1)   & 	23.712503 \\ [1ex]
 \multicolumn{5}{c}{BH/aug-cc-pVQZ}  \\ [1ex]
 Energy &  -25.245606(3)	& 	-25.245609(3)	&	-25.237506(9)	&  	-25.25993075\\
 $\mu_z$ & 0.52656(3)	&	0.52654(5) &	 0.5259(2)	&	0.527200 \\ 
$\alpha_{xx}$	 &	 20.759(1) & 20.761(2) &	20.724(3) &	20.766279 \\
$\alpha_{yy}$	 &	20.759(3)  & 20.758(2) &	20.7234(7) &	20.766279 \\
$\alpha_{zz}$	 &	 23.650(1) & 23.648(2) &	23.625(4) &	23.635628 \\ [1ex]
\hline \hline
\end{tabular*}
\end{center}
\end{table*}

\subsection{Convergence with sampling time}
\label{ssec:rdm-sampling}

In this section we analyze the convergence of the response properties with respect to the number of iterations spent accumulating the transition RDMs between the response and zeroth-order states. We consider the all-electron heteronuclear diatomics LiH and BH as simple test systems, determining all components of the polarizability (spanning totally symmetric and non-symmetric irreps within $C_{2v}$ point group) in \mbox{aug-cc-pV$X$Z} ($X$=D,T,Q) basis sets. All FCIQMC calculations are run with $10^6$ walkers, up to a maximum number of RDM sampling iterations of $10^5$, resulting in small random error bars in this long-time limit. These random errors were found via a blocking analysis of polarizabilities sampled every 10000 iterations from the intermediate tRDMs throughout the simulation, and two replica distributions are used for each state. Fig.~\ref{fig:iter} shows the convergence of these values as the number of sampling iterations ($N_{\rm{iter}}$) is increased at fixed walker number, along with a `best' value from the longest sampling time as a guide for the eye.

Random errors of the polarizability decay with the anticipated $\frac{1}{\sqrt{N_{\rm{iter}}}}$ dependence. However, the size of the random errors increase with basis/system size, reflecting the higher sample variance from this distribution, due to the increased autocorrelation times and smaller timestep required for the same sampling quality. There is also not a large difference between the size of the random errors arising from different components of the perturbation, despite the fact that the $\alpha_{zz}$ estimate sampling was performed via the Hamiltonian spawning steps, while the $\alpha_{yy}$ and $\alpha_{xx}$ components were sampled via the perturbation spawning steps. This is perhaps not too surprising given that these will both have the same number of spawning attempts (and therefore attempted tRDM samples) as there are the same number of walkers in each distribution. The $\alpha_{xx}$ and $\alpha_{yy}$ values should be identical by symmetry arguments, and while their trajectories are distinct, it is also reassuring that we find their values to be in agreement with each other within their random error bars.

\subsection{Convergence with walker number}
\label{ssec:inc_walkers}

In this section, we consider the more important convergence with respect to walker number. While increasing walker number will decrease the random errors in the sampling of expectation values, it is also important to converge the systematic initiator errors which manifest at lower walker numbers. The rate of convergence of this error is key for the success of the approach, as only in the large walker limit does it reach the exact response of the system within the given basis set. The convergence with respect to walker number is therefore a measure of the `initiator' systematic error in the sampling of the zeroth-order and response states, assuming that we are converged with respect to sampling time. 

It has been found previously that this initiator error when applied to energy and property estimators are larger for excited states compared to analogous ground-state expectation values\cite{Blunt2015a, Blunt2017b}. This is because excited states (at least when represented in the canonical Hartree--Fock basis) have a far smaller degree of sparsity in their wavefunction, which makes their discretization in terms of a walker distribution more difficult. Each snapshot in time for a given number of walkers for an excited state will likely give a far worse overall description of the state compared to the analogous ground-state description, and this has ramifications for the severity of the initiator approximation. It is our anticipation that the response states will perform similarly to the convergence of excited states, due to their character as a linear combination of excited state components, but that this can still lead to significant savings compared to many deterministic techniques.

In Fig.~\ref{fig:walker}, we again consider the LiH and BH systems in the \mbox{aug-cc-pV$X$Z} basis sets as in the previous section. Despite being relatively small systems, the Hilbert space size in these large basis sets is over $10^9$ determinants, and therefore for the larger systems we cannot compare to exact FCI response results. Therefore, for a meaningful comparison, we compare to high-level coupled cluster linear response values of the polarizability, with up to full quadruple iterative excitations (CCSDTQ) for the aug-cc-pVDZ basis sets, and CCSDT in the larger basis sets due to computational cost. All coupled cluster calculations for these systems are computed using the General Contraction Code (GeCCo), which consists of a general framework for calculating arbitrary order response properties of coupled cluster methods\cite{Hanauer2009}. Within these coupled cluster results, only the LiH/aug-cc-pVDZ system will be strictly exact and equal to FCI. Furthermore, since the $\alpha_{xx}$ and $\alpha_{yy}$ values are identical in these systems by symmetry, we simply average these values and report the parallel ($\alpha_{||}$) and perpendicular ($\alpha_{\perp}=\alpha_{zz}$) diagonal polarizabilities with respect to the main axis of the molecule. The \mbox{aug-cc-pVDZ} and \mbox{aug-cc-pVTZ} estimators are accumulated for $10^5$ iterations, while for the \mbox{aug-cc-pVQZ} basis they are accumulated for $5\times10^4$ iterations. These are sufficient sampling iterations to converge the random errors to allow for statistically significant deviations from the high-level coupled cluster values to be found.

It should be first noted that for the LiH/aug-cc-pVDZ system where the \mbox{CCSDTQ} benchmark is exact, all components of the polarizability converge exactly to the \mbox{CCSDTQ} values (to within $\sim$0.0002\% of the exact value, and within extremely small random errors), giving confidence in the accuracy of the implementation. While the \mbox{CCSDTQ} value is not necessarily exact for the \mbox{BH/aug-cc-pVDZ} system, it also does not have a statistically significant deviation from the FCIQMC converged values. As the basis sets increase in size, the effect of the initiator approximation and its resultant error at low walker numbers becomes evident. The FCIQMC values requires $\sim10^6$ walkers to converge the aug-cc-pVTZ basis for both systems to within $\sim 10^{-3}$ a.u. of the exact value, and $\sim3\times10^6$ walkers to converge the aug-cc-pVQZ basis for both systems. It also appears that the $\alpha_{||}=\alpha_{zz}$ component converges to the exact value slightly slower than the perpendicular component. This is likely to be due to the fact that the the perpendicular components response wavefunction span a disjoint Hilbert space to the ground-state wavefunction, and therefore does not require the stochastic orthogonalization procedure.

The accuracy of the CCSDT linear response values can also be assessed. These are approximately the same between the aug-cc-pVTZ and aug-cc-pVQZ basis sets, but the errors of the parallel and perpendicular components are very different between the LiH and BH systems. The relative error in the BH system is $\sim 5$ times larger than the LiH system. Furthermore, the parallel and perpendicular component errors are of opposite signs in the BH system, while of the same sign in the LiH system. The relative error of CCSDT compared to the converged FCIQMC value is 0.05\% in each direction in the BH system, while only 0.01\% for the LiH system.


\subsection{Comparison of sampling schemes}
\label{ssec:diff_schemes}

As a final check, we consider the various schemes to obtain averaged response properties from an FCIQMC calculation. Computing averages from random variables requires care, as only averages of linear functions of random variables are free from bias (such as the {\em projected} energy estimator in FCIQMC). To check against non-linear effects biasing the response property estimator, we consider two approaches to average the quantity over the course of the simulation. In the first, we consider the accumulation of the tRDM over the course of the entire simulation, and then a contraction with the perturbation operator at the end in order to obtain the final response value according to Eq.~\ref{eq:pol3}. In order to obtain error bars on this quantity, separate calculations were performed with different random number seeds, and averages and standard errors were obtained between these entirely uncorrelated quantities.

In the second scheme, we consider short periods of accumulation of the tRDMs, before contraction to an intermediate value of the response property on-the-fly. This allows a large number of polarizability estimates to be averaged from a single run. However, the values obtained are serially correlated with each other, requiring a blocking analysis to be performed to obtain an unbiased estimate of the error bar on the averaged value\cite{Flyvbjerg1989}. Despite this, this approach is preferred, since only a single calculation is required, and it is not necessary to perform multiple equilibrations of separate calculations. Since the response property is a linear function of the response vector (see Eq.~\ref{eq:pol2}), this approach should not introduce additional bias into the value obtained.

A final alternative is to only consider a single replica distribution of walkers for each of the zeroth-order and response states. While the bilinear form of the polarizability in Eq.~\ref{eq:pol2} is only explicitly linear in the zeroth-order and response states, these states are in fact correlated with each other through the perturbation spawning and orthogonalization steps seen in Eq.~\ref{eq:ResMaster2}. To remove these correlations, the replica trick is used. However, this increases the computational cost of the calculation by a factor of two, and so we also quantify the bias introduced if only a single replica is used in order to compute the various expectation values. To test these three schemes, we once more consider the LiH and BH systems, and present in Table~\ref{tab:diff_methods} the energy (variational energy from the ground-state density matrix), dipole moment and components of the polarizability tensor. For the systems in aug-cc-pVDZ, aug-cc-pVTZ and aug-cc-pVQZ basis sets, the numbers of walkers used are $1\times10^6$,  $3\times10^6$ and $5\times10^6$, respectively.

From Table~\ref{tab:diff_methods}, we can see that as expected, the averaged values of the polarizability are statistically indistinguishable between those averaged and blocked over the course of a single simulation, and those obtained via the averaging over independent calculations. This validates the simulation protocol used in the previous sections. However, there is a statistically significant bias which can arise in the larger systems when only a single walker replica is used, compared to two independent replicas of each state. This bias is particularly noticable in the largest BH/aug-cc-pVQZ system, where the non-linear bias due to the use of only a single replica gives a relative error of $\sim$0.1-0.2\%. Intriguingly, this polarizability is always underestimated with a single replica. This error is similar in magnitude to that of the non-linear bias in the dipole moment, an expectation value arising from the symmetric RDM, rather than the transition RDM. This indicates that the two replica simulations \emph{are} required for high accuracy calculations of response properties within FCIQMC.

\subsection{Polarizability of CN and NO}
\label{ssec:cn_no}

\begin{figure}[htbp]
\begin{center}
\caption{
Convergence of the initiator errors for parallel ($\alpha_{||}$) and perpendicular ($\alpha_{\perp}$) components of dipole polarizability for CN and NO in a d-aug-cc-pVDZ  basis set as the number of walkers used for sampling the wave functions is increased. 
Results are obtained without correlating $1s$ electrons of C, N and O.
Results are shifted relative to the value, stated here as the best value, corresponding to the highest walker populations involved in the respective studies.
The CCSDT results are obtained from Ref.~\onlinecite{Hammond2008}.
The polarizability component $\alpha_{||}$ is the same as the $\alpha_{zz}$, whereas results for the component $\alpha_{\perp}$ are obtained by averaging over the $\alpha_{xx}$ and $\alpha_{yy}$.
}
\label{fig:CN_NO}
{\includegraphics[trim={0.0cm 0cm 0.0cm 0cm},clip,scale=0.4]{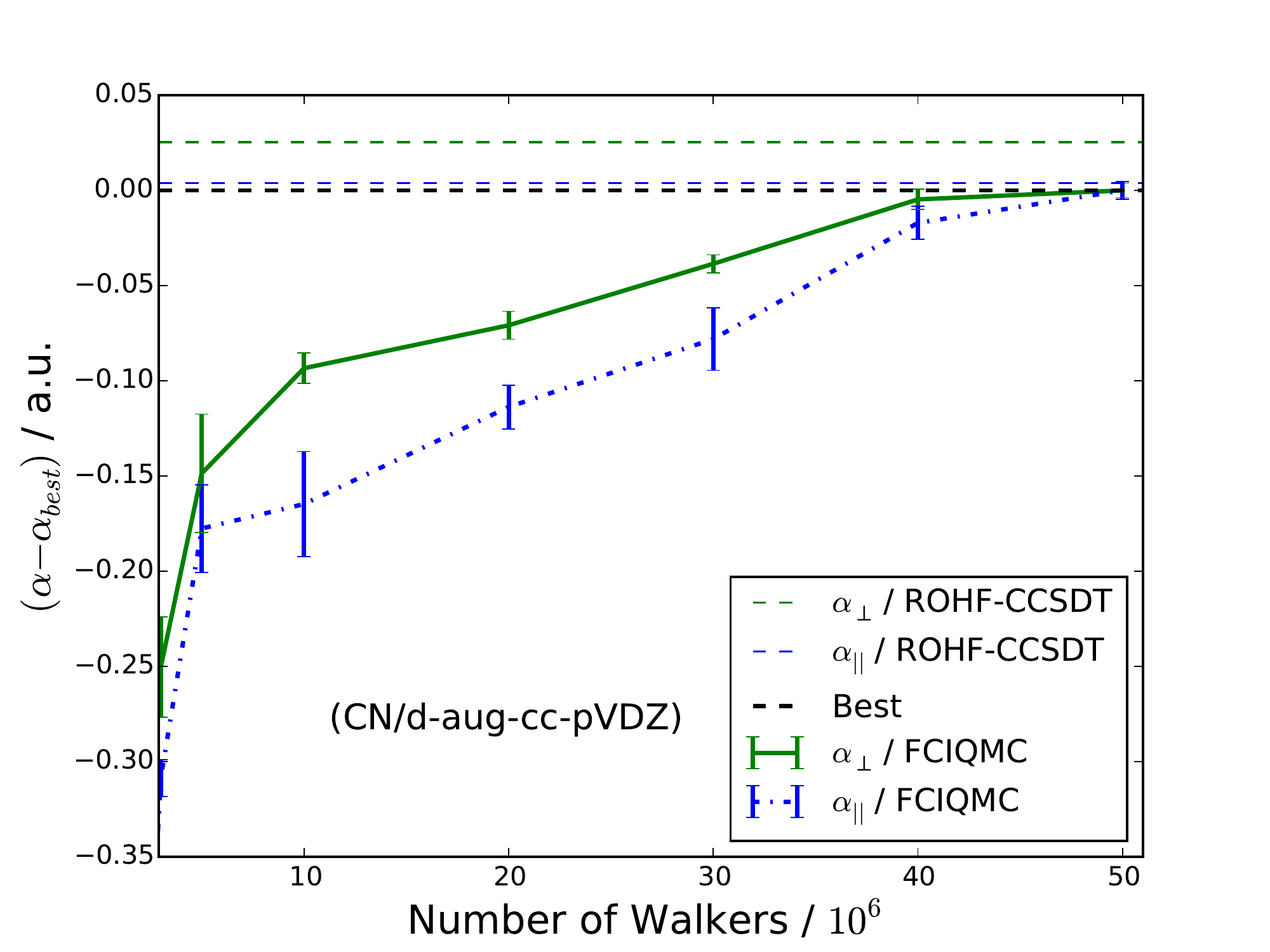}}
{\includegraphics[trim={0.0cm 0cm 0.0cm 0cm},clip,scale=0.4]{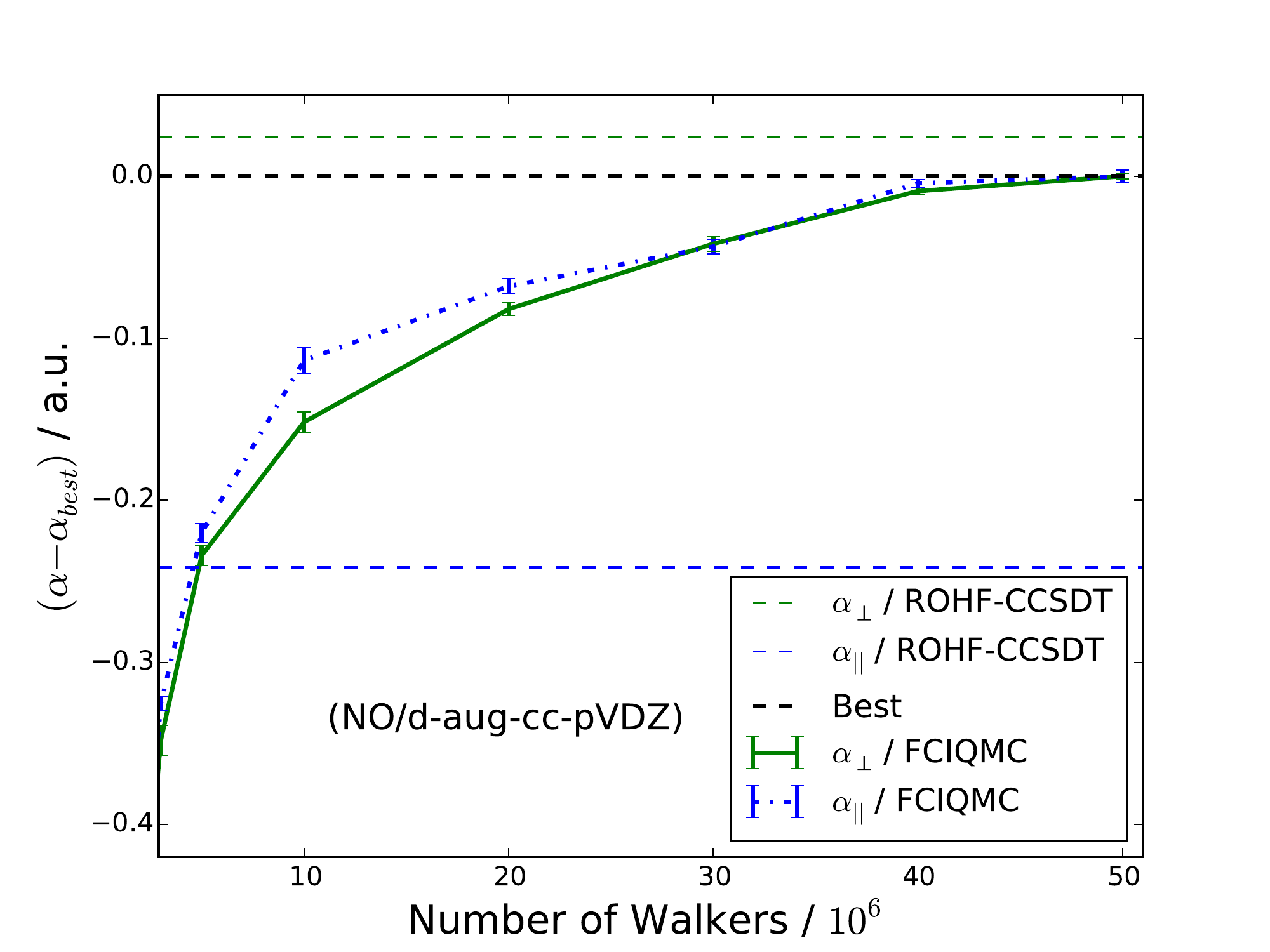}}
\end{center}
\end{figure}

\begin{table*}[htp]
\caption{Parallel ($\alpha_{||}$) and perpendicular ($\alpha_\perp$) components of polarizability in atomic units for the doublet ground states of the CN and NO radicals at a bond length of 1.1718{\AA}, and obtained using d-aug-cc-pVDZ basis set.
Results are obtained without correlating $1s$ electrons of C, N and O.
The final results from FCIQMC are obtained using $5\times10^7$ number of walkers and an ROHF basis. 
}
\begin{tabular*}{\textwidth}{@{\extracolsep{\fill}}lcccccccr} \\ \hline \\ [-2ex]
 &  & \multicolumn{2}{c}{ROHF reference} & \multicolumn{3}{c}{UHF reference}  & & \\ \cline{3-4} \cline{5-7} \\ [-2ex]
Molecule & Components & CCSD\footnotemark[1] & CCSDT\footnotemark[1] & CCSD\footnotemark[2] & CCSDT\footnotemark[2] & CCSDTQ\footnotemark[3] & Experiment\footnotemark[4] &FCIQMC \\ \hline \\ [-1ex]
CN &  $\alpha_{||}$ & 26.398 & 26.349 & 25.587  & 26.267 & 26.432 & & 26.345(5) \\ [0.5ex]
& $\alpha_{\perp}$ & 16.319 & 16.304 & 15.878 & 16.211  &  16.327 & & 16.279(4) \\ [2ex]
NO & $\alpha_{||}$ & 15.546  & 15.408 & 15.521 &  15.406 & & 15.539 &15.649(4)  \\ [0.5ex]
& $\alpha_{\perp}$ & 9.844  & 9.892 & 9.835 &  9.891 &  & 9.844 &9.868(2) \\ [2ex] \hline\hline \\ [-1ex]
\end{tabular*}
\footnotetext[1]{Ref.~\onlinecite{Hammond2008}.} 
\vspace{1em}
\footnotetext[2]{Results for CN and NO are obtained from Ref.~\onlinecite{Kallay2006} and Ref.~\onlinecite{Hammond2008}, respectively.} 
\footnotetext[3]{Ref.~\onlinecite{Kallay2006}.}
\footnotetext[4]{Ref.~\onlinecite{Buckingham1966}.}
\label{tab:CN_NO}
\end{table*}

As a final study, we consider the extension to larger systems, and consider the polarizability of the doublet ground states of the CN and NO radicals, with 13 and 15 electrons respectively and a partially occupied bonding orbital in each. Electrical response quantities are extremely sensitive to both the correlation treatment and basis description of the diffuse, long-ranged part of the orbitals, as these are significantly occupied in the response wavefunction. As a result, we consider the polarizability components in the doubly-augmented d-aug-cc-pVDZ basis, containing 64 orbitals for this system. The values of polarizability are obtained after freezing  $1s$ orbitals of all constituent atoms for both CN and NO.  The Hilbert space for each state, therefore, spans $\sim10^{12}$ functions for the CN, and $\sim10^{14}$ functions for the NO systems -- well beyond traditional FCI treatments.

Hammond \textit{et al.} studied the polarizability of these systems using the linear response formalism for both CCSD and CCSDT methods within a restricted open-shell (ROHF) and unrestricted (UHF) formalism\cite{Hammond2008}.
For CN, additional calculations were also performed at the CCSDTQ level of theory by K\'allay \textit{et al.}, using an unrestricted basis\cite{Kallay2006}. These benchmark studies demonstrated the importance of triple and quadruple excitations in the calculations, but presented an unresolved discrepancy between the UHF and ROHF values at the CCSDT level which are in significant disagreement for the CN radical, but relatively similar for the NO system. 

We use two replicas for each state, and an ROHF basis for our FCIQMC simulations, though this choice should not change results since FCIQMC at convergence is invariant to the choice of basis (as the complete space is always spanned) and all states should be free from spin-contamination.
The components of the polarizability tensor are presented as $\alpha_{||}=\alpha_{zz}$ and $\alpha_{\perp}=\frac{1}{2}(\alpha_{xx}+\alpha_{yy})$, with \textit{z} being the bond-axis, at a bond length of 1.1718{\AA}.
Convergence of $\alpha_{||}$ and $\alpha_{\perp}$ for both CN and NO with an increasing number of walkers is shown in Fig.~\ref{fig:CN_NO}. We estimate that by 50 million walkers, we have converged all polarizability components for both systems to $\sim$0.01-0.02 a.u., or between 0.05-0.1\% error in the polarizability components. This allows for statistically significant determination of the errors in the coupled-cluster linear response values. While the absolute convergence of these response quantities is relatively slow compared to quantities which depend solely on the ground-state walker distribution (such as the energy and dipole moment), the rate of convergence is similar. The analogous walker convergence of these systems for the ground-state energy and dipole moment are shown explicitly in the Supplementary Information for comparison\cite{SuppInfo}. 

The CCSDT ROHF-reference linear response polarizability values are also shown for comparison in Fig.~\ref{fig:CN_NO} and are generally in good agreement with the FCIQMC values, with the notable exception of the parallel component of the NO radical, which is in error by $\sim$1.5\%. These values are further analyzed, and compared to their UHF counterparts in Table~\ref{tab:CN_NO}. This error in $\alpha_{||}$ for NO arises despite good agreement between the UHF and ROHF CCSDT values, and good agreement with FCIQMC for $\alpha_{\perp}$. While not conclusive, further evidence of the error comes from comparison to experimental results, which are in good agreement with FCIQMC, and the $\alpha_{\perp}$ CCSDT values, but again in significant error for $\alpha_{||}$ where it appears that beyond-triples contributions are required, regardless of reference state. However, full agreement between FCIQMC and experiment would only be expected in the complete basis set limit, and with appropriate consideration of vibrational corrections. 

For the CN system, we find that the ROHF-CCSDT linear response values are in good agreement with FCIQMC, however the UHF-based values are also in significant error, which are not corrected even via the inclusion of quadruple excitations. Indeed, the ROHF-CCSDT values for $\alpha_{||}$ and $\alpha_{\perp}$ are found to be substantially more accurate compared to FCIQMC than both the UHF-CCSDT and UHF-CCSDTQ values. This indicates that the error introduced by the additional spin contamination is too severe for the UHF-based coupled cluster results for the polarizability of this system to be corrected even by the inclusion of quadruple excitations.



\section{Conclusions and future work}
\label{sec:}

In this paper, we have presented a general formulation of arbitrary-order response theory for FCIQMC. 
We limited the scope of our numerical study to static, linear-response properties, and therefore the sampling of the frequency-independent 
first-order response vector.
This response vector was sampled via a modified set of stochastic master equations which aimed to solve a coupled set of linear equations rather than the traditional zeroth-order state eigenvalue problem. It was necessary for the zeroth-order state walker distribution to evolve stochastically alongside the walkers representing the response function, and the two walker distributions were coupled via a spawning dynamic from the zeroth-order state walkers into the response.
In an unbiased fashion, we found that it was possible to stochastically sample both these two states, and the resultant response properties via a contraction of the transition RDM, using the replica approach that has been developed previously for first-order properties\cite{Blunt2017b}. The initiator criterion was trivially applied to the sampling of the response dynamics, which led to a convergence of the response property with respect to increasing walker number towards the exact response properties.

We demonstrated the approach by computing the static electric dipole polarizability for a set of diatomic molecules, taking care to consider convergence with sampling time, walker number and any potential non-linear bias which may result from the sampling. We moved to the larger NO and CN radical systems, where stronger correlation effects have shown a requirement of triple and quadruple excitations to compute the polarizability. Converging these values with walker number, we were able to shed light on the discrepancies between the UHF and ROHF values for these systems when comparing to the previous existing best estimates, obtained from high-level coupled cluster up to full quadruple excitations. From this, we found the UHF-based values to be significantly in error compared to their ROHF-based counterparts.

Although applied only to relatively small systems in this work, this response theory within the FCIQMC stochastic framework paves the way to use FCIQMC for calculating a variety of linear as well as higher-order response properties with very high accuracy, as well as an extension to dynamic quantities.
It will also be important to investigate schemes to improve upon the rate of convergence of these response functions as the walker number is increased. This will consider the use of optimized basis sets to maximize sparsity in the response wavefunction\cite{Thomas14}, as well as relaxed density matrices to take into account frozen core electrons or an active-space construction\cite{Thomas15}, and investigations into additional approximations and modifications of the stochastic rules to accelerate convergence. These have recently been formulated for the ground state problem via perturbative corrections and size-consistency modifications to the algorithms\cite{Tenno17,Blunt2018}. Finally, it will be considered whether a similar approach for non-linear parameterizations of wavefunction forms will also be amenable within this framework\cite{Schwarz17}.

\section{Acknowledgments}
\label{sec:acknowledgments}

G.H.B. gratefully acknowledges funding from the Air Force Office of Scientific Research via grant number FA9550-16-1-0256, and the Royal Society via a University Research Fellowship. N.S.B. gratefully acknowledges St. John's College, Cambridge for funding through a Research Fellowship. We are also grateful to the UK Materials and Molecular Modelling Hub for computational resources, which is partially funded by EPSRC (EP/P020194/1). 
P.K.S gratefully acknowledges Graduate School Sim Tech at the University of Stuttgart for funding an extended collaborative visit to King's College London where the work was carried out. P.K.S. thanks Robert Anderson for fruitful discussions and helpful advice about the code, and Andreas K\"ohn for helpful discussions. 

\end{document}